\documentclass[aps,prl,reprint,superscriptaddress,nofootinbib,floatfix,nobibnotes]{revtex4-2}
\usepackage[T1]{fontenc}
\usepackage{mathptmx}
\usepackage{amsmath,amssymb,bm,graphicx,xcolor}
\usepackage[colorlinks=true,linkcolor=blue,citecolor=blue,urlcolor=blue]{hyperref}
\graphicspath{{figures/}}

\newcommand{\Tc}[1]{#1\,^{\circ}\mathrm C}

\hypersetup{
  pdftitle={Recoverable Necking Enabled by Reverse Phase-Boundary Motion in Liquid Crystal Elastomers},
  pdfauthor={Weida Kang, Yufeng Zhang, Changyue Liu, Zhijian Wang, Xudong Liang}
}
\DeclareMathAlphabet{\mathcal}{OMS}{cmsy}{m}{n}
\makeatletter

\def\LCE@printnote#1{%
  \def\@TBN@opr##1##2{%
    \ifnum##1=#1\relax
      \present@FM@footnote{##1}{##2}%
    \fi
  }%
  \@FMN@list
}

\def\frontmatter@footnote@produce{%
  \LCE@printnote{1}%
  \LCE@printnote{3}%
  \LCE@printnote{2}%
  \global\let\@FMN@list\@empty
}

\makeatother

\begin{document}

\title{Recoverable Necking Enabled by Reverse Phase-Boundary Motion in Liquid Crystal Elastomers}

\author{Weida Kang}
\thanks{These authors contributed equally to this work.}
\affiliation{School of Science, Harbin Institute of Technology, Shenzhen, 518055, China}

\author{Yufeng Zhang}
\thanks{These authors contributed equally to this work.}
\affiliation{School of Science, Harbin Institute of Technology, Shenzhen, 518055, China}

\author{Changyue Liu}
\thanks{These authors contributed equally to this work.}
\affiliation{Department of Mechanical and Automation Engineering, The Chinese University of Hong Kong, Hong Kong, 999077, China.}
\affiliation{Key Laboratory of Aerospace Advanced Materials and Performance, Ministry of Education, School of Materials Science and Engineering, Beihang University, Beijing, 100191, China.}

\author{Zhijian Wang}
\email[Contact author: ]{zhijianw@buaa.edu.cn}
\affiliation{Key Laboratory of Aerospace Advanced Materials and Performance, Ministry of Education, School of Materials Science and Engineering, Beihang University, Beijing, 100191, China.}
\affiliation{Tianmushan Laboratory, Hangzhou, 310023, China.}

\author{Xudong Liang}
\email[Contact author: ]{liangxudong@hit.edu.cn}
\affiliation{School of Science, Harbin Institute of Technology, Shenzhen, 518055, China}

\begin{abstract}
Strain localization is often associated with plastic flow and permanent deformation. Here we show that a tensile neck in a crosslinked nematic network can undergo nearly complete recovery through reverse phase-boundary motion, leaving little residual deformation. In a monodomain liquid crystal elastomer stretched perpendicular to its initial director, this recovery occurs during an isothermal loading-unloading cycle. Digital image correlation resolves the process as the retreat of interfaces separating low- and high-stretch regions, where the high-stretch domain shrinks while previously necked material thickens, despite substantial mechanical hysteresis. Within the recoverable-necking regime, increasing temperature lowers the loading plateau stress but raises the unloading plateau stress. A nonconvex constitutive description provides an energetic basis for the coexistence of the two stretch states, while gradient regularization accounts for the finite width of their interfaces through an equilibrium reference profile. These results identify the spatial mechanism of neck recovery and distinguish recoverable strain localization from mechanical reversibility in an orientationally ordered polymer network.

\end{abstract}
\maketitle
Strain localization concentrates deformation into a small region, but does not by itself determine whether the
deformation is recoverable. In conventional metals, tensile necking accompanies plastic flow and can precede
failure, while yielding and strain softening in glassy polymers can leave permanent deformation~\cite{HawardThackray1968}. Nonlinear elasticity also permits localization through the coexistence of distinct deformation states\cite{AudolyHutchinson2016,LestringantAudoly2020,GiudiciBiggins2020}. Liquid crystal elastomers (LCEs), in which network deformation couples to internal orientational order \cite{Warner2003}, provide a setting for exploring recoverable strain localization.

Necking and nonuniform deformation in nematic elastomers have been studied under different director orientations and mechanical boundary conditions~\cite{Higaki2013,Poudel2024,Kang2025JMPS,Chehade2024,Poudel2026}. He et al.~\cite{He2020} reported neck recovery for oblique initial director orientations, while Ohzono et al.~\cite{Ohzono2021} showed that
nematic-order constraints can retard or arrest necking recovery. These findings demonstrate that necking recovery in nematic elastomers depends sensitively on molecular orientation and network constraints. Yet, the spatial mechanism by which an established neck disappears, and its connection to interface propagation and mechanical hysteresis, remain unclear.

Here we resolve the spatial mechanism of recoverable necking in a monodomain LCE stretched perpendicular to its initial director. Digital image correlation (DIC) reveals that unloading erases the neck through the retreat of the interfaces bounding the high-stretch domain, accompanied by local thickening of the previously necked material. By tracking these interfaces over a range of temperatures, we identify distinct temperature-dependent propagation stresses for forward and reverse conversion. A gradient-regularized free-energy framework further connects the coexistence of deformation states to the finite-width interfaces and the energetic accessibility of reverse localization. These results establish reverse phase-boundary motion as a mechanism for recovering localized deformation without requiring a reversible mechanical cycle.

\begin{figure*}[t]
\centering
\includegraphics[width=\textwidth]{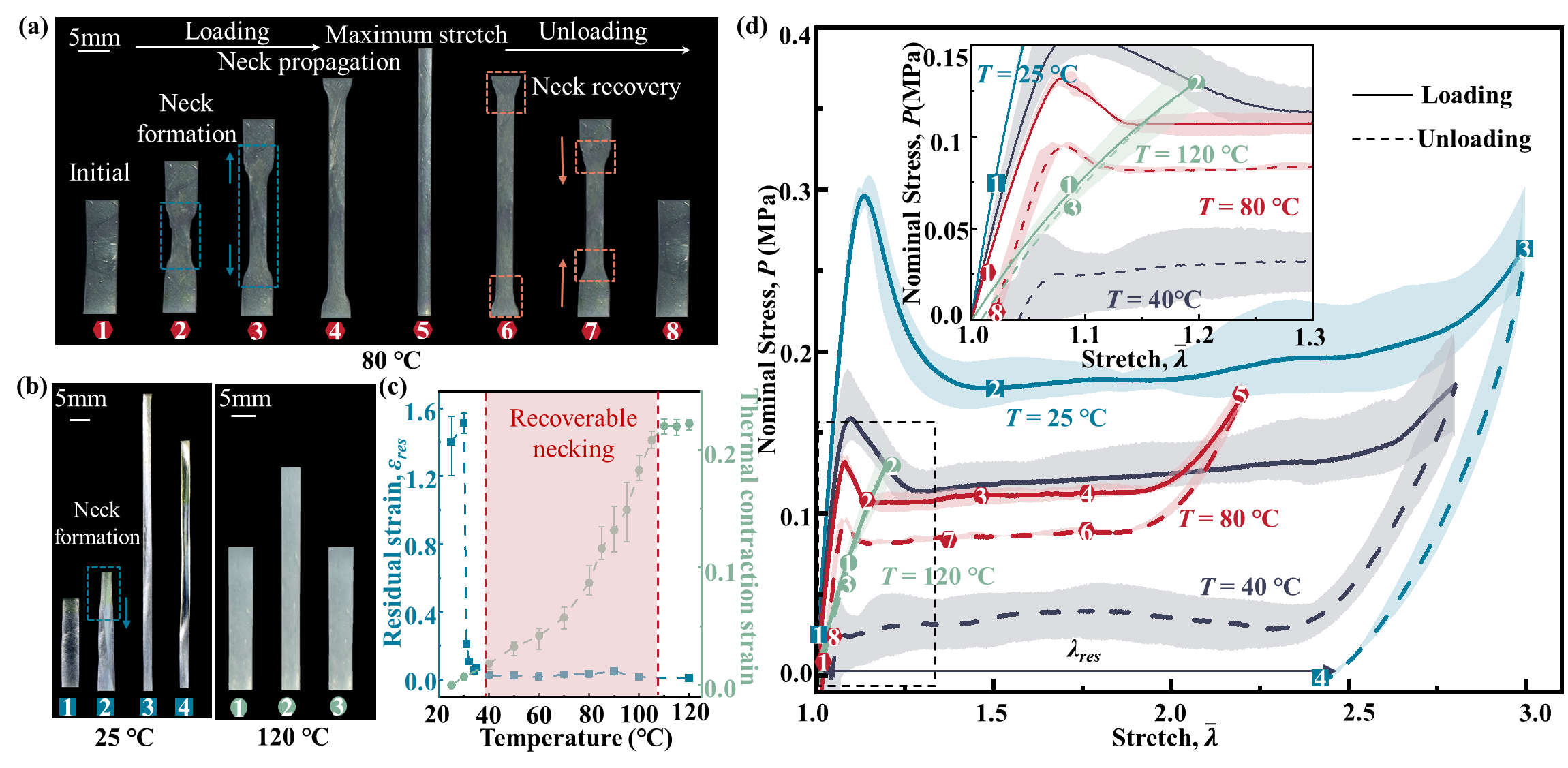}
\caption{Temperature-dependent necking and recovery responses. (a) Loading and unloading at $\Tc{80}$. (b) Configurations at $\Tc{25}$ and $\Tc{120}$. (c) Residual strain after unloading and thermal-contraction strains at different temperatures. Shading marks the observed recoverable-necking regime. (d) Nominal stress-stretch curves. Solid and dashed curves show the mean loading and unloading responses, respectively. Shading spans the minimum and maximum of three independent tests. Numbered markers correspond to the configurations in (a) and (b). The inset magnifies the small-stretch responses.}
\label{fig:one}
\end{figure*}

Neck formation and shape recovery exhibit distinct temperature dependence. Each specimen was first allowed to reach thermal equilibrium in its stress-free configuration and was then loaded and unloaded isothermally. At \(80\,^\circ\mathrm{C}\), the necked region grows during loading and retreats during unloading, while the previously narrowed material widens again [Fig.~\ref{fig:one}(a)]. At room temperature of \(25\,^\circ\mathrm{C}\), necking leaves substantial residual deformation after unloading. At \(120\,^\circ\mathrm{C}\), above the transition range characterized in the Supplemental Material (SM)~\cite{SM}, the specimen recovers without pronounced necking over the tested stretch range [Fig. 1(b)]. These contrasting responses are quantified by the residual strain \(\varepsilon_{\mathrm{res}}=\lambda_{\mathrm{res}}-1\), which decreases sharply with temperature [Fig.~\ref{fig:one}(c)]. The shaded region identifies a temperature window above room temperature but below the nematic-isotropic transition temperature ($T_{NI} \simeq 110^\circ\mathrm{C}$, Fig.~S2(c) in the SM~\cite{SM}) in which a pronounced neck forms and subsequently disappears with nearly complete macroscopic shape recovery. The free thermal contraction strain plotted alongside the residual strain characterizes the change of the stress-free shape upon heating [Fig. 1(c)]. Neck recovery, by comparison, occurs during unloading within a temperature window, as illustrated by Movie S1 in the SM~\cite{SM}.

To compare the mechanical responses across temperatures, we define the mean stretch and nominal stress as \(\bar{\lambda}=L/L_g(T)\) and \(P=F/A_0(T)\), where \(F\) is the axial force, and \(L_g(T)\) and \(A_0(T)\) are the stress-free gauge length and cross-sectional area at the testing temperature. At \(80\,^\circ\mathrm{C}\), a stress maximum is followed by a broad loading plateau as the neck grows, whereas unloading follows a lower plateau as the neck retreats [Fig.~\ref{fig:one}(d)]. Within the recoverable-necking regime, lowering temperature increases the loading plateau stress but reduces the unloading plateau stress, enhancing the separation between the two paths (gray curve at \(40\,^\circ\mathrm{C}\)). At \(120\,^\circ\mathrm{C}\), the loading and unloading curves nearly coincide and show no observable plateau. Near-complete recovery of the neck within the nematic regime thus occurs while the mechanical paths remain distinct. Therefore, the localized deformation can recover even though the loading-unloading cycle retains substantial hysteresis. 

\begin{figure}[t]
\centering
\includegraphics[width=\columnwidth]{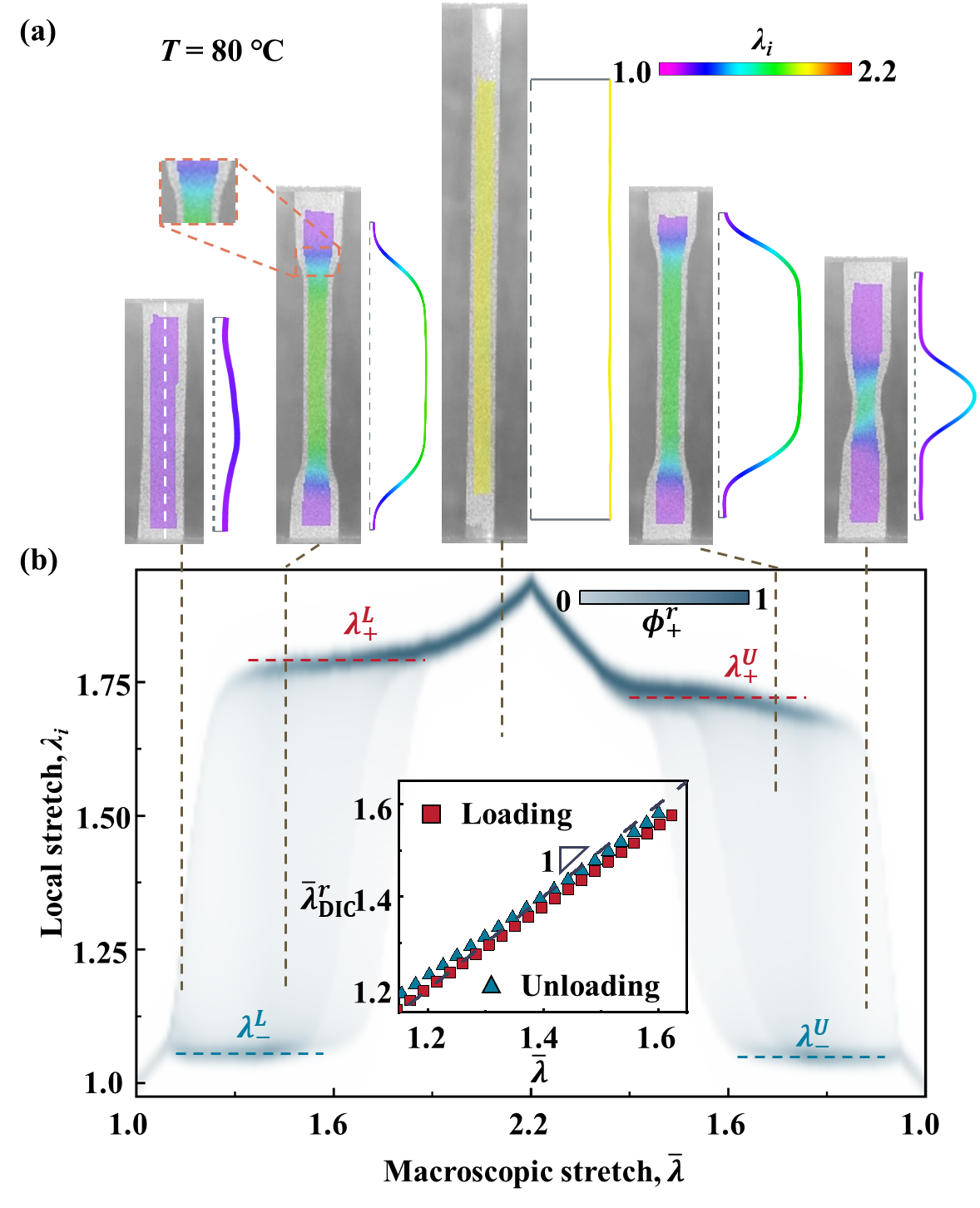}
\caption{DIC-resolved phase conversion at $\Tc{80}$. (a) Local stretch fields $\lambda_i$ and corresponding centerline profiles during loading and unloading. Enlarged view of an interface shown in the dashed-box. (b) Local stretch evolution during loading and unloading. Darker colors indicate a larger fraction. Horizontal dashed lines mark the low- and high-stretch levels $\lambda_{\pm}^r$. Inset: mean stretch reconstructed using the DIC-based lever rule($\lambda_{\rm DIC}^r$) versus the imposed mean stretch $\bar \lambda$.}
\label{fig:two}
\end{figure}

DIC at \(80\,^\circ\mathrm{C}\) reveals the spatial pathway of neck recovery [Fig.~\ref{fig:two}(a)]. During loading, the initially nearly uniform stretch field develops low- and high-stretch regions connected by finite-width interfaces. On unloading, these interfaces retreat and the high-stretch region shrinks, returning previously necked material toward the low-stretch state. Near-incompressibility links this local reduction in axial stretch to an increase in cross-sectional area upon unloading, consistent with the observed widening of the previously necked materials in Fig.~\ref{fig:one}(a). Movie S2 in the SM~\cite{SM} shows the evolution of the local stretch fields at different temperatures.

Over most of the coexistence interval, the bulk stretch levels \(\lambda_-^r\) and \(\lambda_+^r\) vary relatively little along each path, whereas the relative fractions of the two regions change substantially [Fig.~\ref{fig:two}(b)]. Here, $r=L,U$ denotes loading and unloading, respectively. Approximating the measured field by two uniform stretch regions gives the length-compatibility relation
\begin{equation}
\bar{\lambda}_{\mathrm{DIC}}^r
\simeq
(1-\phi_+^r)\lambda_-^r+\phi_+^r\lambda_+^r,
\label{lever-rule}
\end{equation}
where \(\phi_+^r\) denotes the high-stretch fraction in the reference configuration. The reconstructed stretch closely follows the imposed mean stretch over the range shown in the inset of Fig.~\ref{fig:two}(b), consistent with extension and recovery being accommodated primarily by changes in the relative fractions of the two stretch states. The high-stretch plateau is lower during unloading than during loading, \(\lambda_+^U<\lambda_+^L \), demonstrating that the reverse transformation follows a distinct local deformation path. These observations identify neck recovery as the retreat of a localized high-stretch domain and show that near-complete macroscopic recovery does not require retracing the same local stretch states.

\begin{figure*}[t]
\centering
\includegraphics[width=\textwidth]{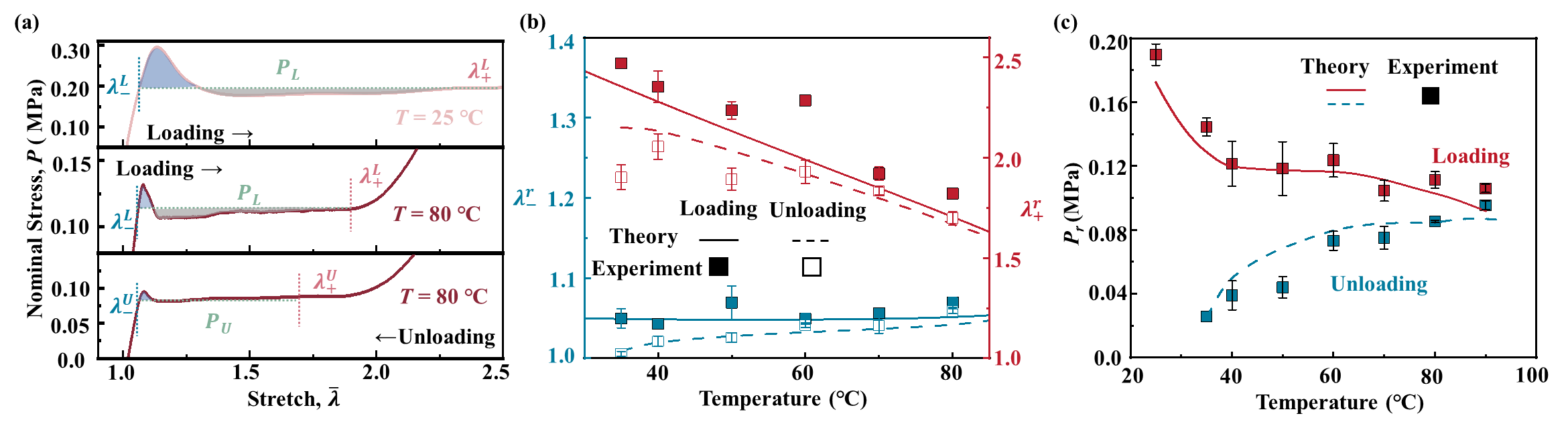}
\caption{Phase selection in necking. (a) Experimental equal-area constructions at $\Tc{25}$ and $\Tc{80}$. (b,c) Coexisting stretch levels and propagation stresses for loading and unloading. Symbols denote experimental measurements. Solid curves show the loading predictions from the calibrated Maxwell construction, while dashed curves show the corresponding unloading results, with the local stretch states evaluated at the measured $P_U(T)$ using Eq.~(3). Error bars indicate the standard deviation of three independent tests.}
\label{fig:three}
\end{figure*}

To connect the observed stretch domains to their energetics, we reduce the three-dimensional constitutive free energy to uniaxial extension. The energy combines network elasticity, orientational memory, and nematic ordering. Its density per unit volume of the stress-free reference configuration at the testing temperature is denoted by \(\Psi_T(\mathbf F_m,\bm{d},Q;T)\), consistent with the stretch and stress measures used above. For extension along \(\mathbf e_z\), perpendicular to the initial director \(\mathbf e_y\), we restrict the homogeneous deformation to \(\mathbf F_m=\operatorname{diag}[\lambda_x,(\lambda\lambda_x)^{-1},\lambda]\) and \(\bm{d}(\theta)=\cos\theta\,\mathbf e_y+\sin\theta\,\mathbf e_z\). Minimizing over the transverse stretch \(\lambda_x\), director angle \(\theta\), and scalar order parameter \(Q\) at fixed axial stretch gives
\begin{equation}
\Phi(\lambda,T)
=
\min_{\lambda_x,\theta,Q}
\Psi_T\!\left(
\mathbf F_m(\lambda,\lambda_x),
\bm{d}(\theta),Q;T
\right).
\label{free energy}
\end{equation}
At intermediate director orientations, axial extension is generally coupled to shear. Here, we adopt a diagonal approximation for \(\mathbf F_m\), restricting the admissible kinematics to axial and transverse stretches. This approximation is motivated by the small shear strain relative to the axial strain observed in experiments (Fig.~S6 in the SM~\cite{SM}). The constitutive energy and its reduction are presented in Appendix~\ref{app:relaxation}, with calibration details provided in the SM~\cite{SM}. For the calibrated parameters, the resulting reduced energy \(\Phi\) is nonconvex in the necking regime (Fig.~S7 in the SM~\cite{SM}), allowing coexistence of low- and high-stretch regions to lower the bulk free energy at fixed mean stretch.

We determine the experimentally resolved propagation stresses \(P_ L(T)\) and \(P_U(T)\) from equal-area constructions over the coexistence intervals defined by the DIC-resolved stretch levels $\lambda_-^r$ and $\lambda_+^r$ ($r=L,U$). Fig.~\ref{fig:three}(a) illustrates the constructions for loading at room temperature and for both loading and unloading at $\Tc{80}$. These experimental plateau estimates are distinguished from the equilibrium coexistence stress of the model. Introducing the Gibbs free-energy density \(g(\lambda;P,T)=\Phi(\lambda,T)-P\lambda\), the two locally stable stretch states satisfy
\begin{equation}
\Phi'(\lambda_\pm,T)=P,
\qquad
\Phi''(\lambda_\pm,T)>0,
\end{equation}
where primes denote derivatives with respect to \(\lambda\) at fixed temperature. Their energy-density difference is \(\Delta g(P,T)=g(\lambda_+;P,T)-g(\lambda_-;P,T)\).

The common tangent to the loading-calibrated energy defines the effective Maxwell stress $P_M^{\mathrm{eff}}(T)$ and the loading coexistence stretches. The unloading stretch levels are obtained from the stable roots of Eq.~(3) at the measured stress $P_U(T)$, capturing the experimental trends in Fig.~\ref{fig:three}(b). Evaluating the Gibbs-energy difference at $P_U$ defines $\Delta g_c(T)=\Delta g(P_U(T),T)$, which is positive for $P_U<P_M^{\mathrm{eff}}$ and favors conversion from the high- to the low-stretch state.

Fig.~\ref{fig:three}(c) reveals opposite temperature trends of the forward and reverse propagation stresses. The loading-calibrated Maxwell stress decreases, whereas the measured unloading stress increases. Integrating $d\Delta g/dP=-\Delta\lambda(P,T)$ along the stable branches and approximating the stretch contrast by its Maxwell value (Appendix~\ref{app:coexistence}) gives
\begin{equation}
P_M^{\mathrm{eff}}(T)-P_U(T)
\simeq
\frac{\Delta g_c(T)}{\Delta\lambda_M(T)}, 
\label{eq:plateau_gap}
\end{equation}
where \(\Delta\lambda_M(T)=\lambda_+(P_M^{\mathrm{eff}},T)-\lambda_-(P_M^{\mathrm{eff}},T)\) is the stretch contrast at the Maxwell reference. The resulting decrease in $\Delta g_c$ indicates a reduced bulk energetic bias against reverse conversion at higher temperatures.

\begin{figure*}[t]
\centering
\includegraphics[width=\textwidth]{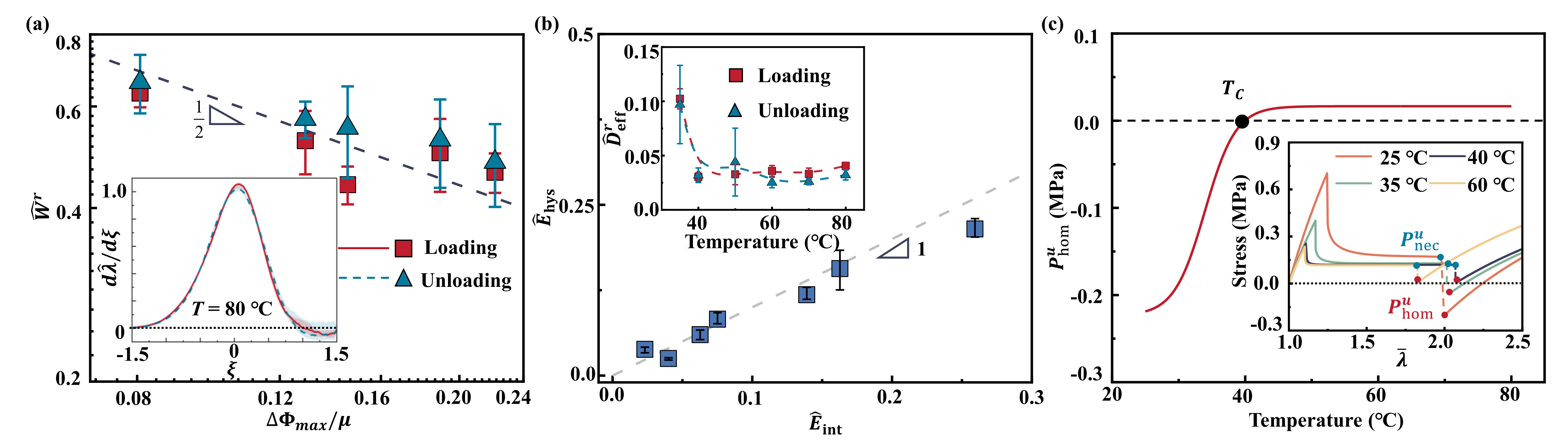}
\caption{Interface structure and finite-length energetics. (a) Necking width scaling with energy barrier between two phases. Symbols show width measured from DIC results. Inset: normalized necking profiles in loading and unloading at $\Tc{80}$. (b) Comparison between the hysteresis work and the sum of interfacial energy for a loading-unloading cycle. Inset: Experiment-calibrated dimensionless gradient coefficients. (c) Temperature dependence of the homogeneous-branch nominal stress $P_{\mathrm{hom}}^u$ at the high-stretch equal-energy crossing. The black marker identifies $T_c$, where $P_{\mathrm{hom}}^u=0$. Inset: nominal stress--stretch curves of the homogeneous and localized states at $\Tc{25}$, $\Tc{35}$, $\Tc{40}$, and $\Tc{60}$. Paired markers indicate the two branch stresses, $P_{\mathrm{hom}}^u$ and $P_{\mathrm{nec}}^u$, at the same mean stretch $\bar\lambda_u(T)$.}
\label{fig:four}
\end{figure*}

The bulk energy selects the coexisting stretch states but provides no intrinsic interfacial length scale. At $80\,^\circ\mathrm{C}$, normalized loading and unloading profiles nearly superpose despite their distinct propagation stresses [inset of Fig.~\ref{fig:four}(a)]. Here, $\hat{\lambda}=(\lambda-\lambda_-^r)/\Delta\lambda^r$ and $\xi=s(Z-Z_f)/W^r$, where $r=L,U$ labels the loading and unloading paths, $\Delta\lambda^r=\lambda_+^r-\lambda_-^r$, $Z_f$ and $W^r$ denote the interface midpoint and width, and $s=\pm1$ orients each profile from low to high stretch (see SM~\cite{SM}). To describe these diffuse interfaces, we introduce (Appendix~\ref{app:gradient})
\begin{equation}
\mathcal{F}[\lambda]=\int_0^{L_g}
\left[
A\Phi(\lambda,T)
+\frac{D_{\mathrm{eff}}}{2}(\lambda_{,Z})^2
\right]\mathrm{d}Z,
\label{eq:gradient}
\end{equation}
where the effective gradient coefficient $D_{\mathrm{eff}}>0$ characterizes the energetic cost of spatial variations in stretch \cite{AudolyHutchinson2016,Yu2025JMPS} under the adopted kinematic description (see SM~\cite{SM}). 
At each temperature, we approximate the gradient coefficient as uniform across the interface	and	estimate it from the measured stretch profiles from DIC.

For an isolated stationary interface at the loading-calibrated Maxwell reference, the equilibrium equation admits the first integral (Appendix~\ref{app:gradient})
\begin{equation}
\frac{D_{\mathrm{eff}}}{2A}(\lambda_{,Z})^2
=\Delta\Phi(\lambda,T).
\label{eq:first}
\end{equation}
Here, $\Delta\Phi$ is the excess above the common tangent connecting the loading coexistence stretches $\lambda_\pm^L(T)$ and vanishes at both endpoints (see SM~\cite{SM}). Separate fits of the loading and unloading profiles to this common reference yield $D_{\mathrm{eff}}^r(T)$ of comparable magnitude [inset of Fig.~\ref{fig:four}(b)]. For unloading, the reference characterizes the interface profile, while propagation occurs at the measured stress $P_U(T)$.

The balance between bulk and gradient energies sets the equilibrium interfacial width,
\begin{equation}
\hat W^r \simeq
C_W(T)\frac{\Delta\lambda_M}{\Delta\lambda^r}
\sqrt{\hat D_{\mathrm{eff}}^r}
\left(
\frac{\Delta\Phi_{\max}}{\mu}
\right)^{-1/2},
\label{eq:width_scaling}
\end{equation}
where $\hat W^r=W^r/(b\Delta\lambda^r)$, $\hat D_{\mathrm{eff}}^r=D_{\mathrm{eff}}^r/(A\mu b^2)$, $\Delta\lambda_M=\lambda_+^M-\lambda_-^M$, and $b$ is the reference specimen width. The factor $C_W(T)$ depends on the normalized barrier shape and the width convention. The width scaling follows from the balance between bulk and gradient energies. Fig.~\ref{fig:four}(a) shows that both loading and unloading widths broadly follow the inverse-square-root scaling. The reference exponent assumes limited variation of $\hat{D}_{\mathrm{eff}}^r$ and $C_W$. 

The equilibrium interface profile also determines the excess free energy per unit reference area. For the Maxwell profile, the bulk excess and gradient contributions are equal, giving the interfacial excess energy per unit reference area (Appendix~\ref{app:gradient})
\begin{equation}
\gamma_{\mathrm{int}}^r(T)
=
\frac{1}{A}
\int_{\lambda_-^M}^{\lambda_+^M}
\sqrt{
2AD_{\mathrm{eff}}^r(T)\Delta\Phi(\lambda,T)
}
\,\mathrm d\lambda.
\label{eq:interface_energy}
\end{equation}
The boundary therefore stores energy both in intermediate stretch states and in spatial gradients. We compare this interfacial energy scale with the measured hysteresis work using
\begin{equation}
\begin{aligned}
\hat E_{\mathrm{int}}
&=
\frac{N_{\mathrm{int}}(\gamma_{\mathrm{int}}^L+\gamma_{\mathrm{int}}^U)}{\mu L_g},\\
\hat E_{\mathrm{hys}}&=\frac{1}{\mu}\left(\int_{L}P\,\mathrm d\bar\lambda+\int_{U}P\,\mathrm d\bar\lambda
\right)
\end{aligned}
\label{eq:energy_comparison}
\end{equation}
where $N_{\mathrm{int}}=2$ for a neck bounded by two fronts. The interface stores excess free energy in both intermediate stretch states and spatial gradients. Fig.~\ref{fig:four}(b) compares the path-summed reference interface energy with the measured mechanical hysteresis. Their comparable magnitudes show that interfacial costs are appreciable on the scale of the mechanical cycle. However, interface translation need not accumulate proportional interfacial energy, and cycle work equals dissipated energy only when the complete thermodynamic state returns to its initial value.

For a finite specimen, localization can lower the bulk free energy through stretch redistribution but incurs
an interfacial cost. We compare the total Helmholtz free energies of the homogeneous and localized configurations, $\mathcal F_{\mathrm{hom}}=AL_g\Phi$ and $\mathcal F_{\mathrm{nec}}=\mathcal F[\lambda_i;T]$, respectively, where $\lambda_i$ is a nonuniform stationary profile at prescribed mean stretch. Their energy-density difference is $\Delta f=(\mathcal F_{\mathrm{nec}}
-\mathcal F_{\mathrm{hom}})/(AL_g)$.

The configurational equal-energy condition requires $\Delta f(\bar\lambda_u,T)=0$, where $\bar\lambda_u$ is the critical stretch upon unloading. Along either stationary configuration, $\mathrm d\mathcal F_\alpha/\mathrm d\bar\lambda=AL_gP_\alpha$, where $\alpha=\mathrm{hom},\mathrm{nec}$, so that $\partial\Delta f/\partial\bar\lambda
=P_{\mathrm{nec}}-P_{\mathrm{hom}}$. At the high-stretch equal-energy crossing, the two configuration stresses generally differ [Inset in Fig.~\ref{fig:four}(c) for different temperatures.]. For a small contraction in unloading, $P_{\mathrm{nec}}^u>P_{\mathrm{hom}}^u$ makes the localized deformation energetically preferred (Appendix~\ref{app:finite-length}). 

Fig.~\ref{fig:four}(c) plots $P_{\mathrm{hom}}^u(T)$ at this equal-energy crossing. Within the idealized equal-energy switching description, the zero-load boundary is $P_{\mathrm{hom}}^u(T_c)=0$, with $T_c\approx40\,^\circ\mathrm C$. For $P_{\mathrm{hom}}^u>0$, the crossing is reached under tension, making reverse localization energetically accessible before the zero-force endpoint. For $P_{\mathrm{hom}}^u<0$, the homogeneous branch reaches zero force first and can retain substantial residual stretch. Within the observed temperature window between $T_c$ and $T_{\mathrm{NI}}$, reverse conversion produces nearly complete macroscopic shape recovery despite finite hysteresis.

In summary, we identify reverse boundary motion as the mechanism underlying recoverable necking in a nematic elastomer. During unloading, the high-stretch domain contracts and disappears, restoring the macroscopic shape despite substantial hysteresis. Nonconvex bulk elasticity selects the competing stretch states, while gradient regularization determines the finite width and energetic cost of their interfaces. Together with the temperature-dependent propagation stresses and finite-length energy comparison, these results connect interfacial energetics to the accessibility of reverse localization. They demonstrate that recoverable strain localization does not require a mechanically reversible cycle, but can emerge through the energetics and motion of internal boundaries. This mechanism provides a broader perspective on deformation recovery in nonlinear elastic materials with internal degrees of freedom.

\par\smallskip
\vspace{\baselineskip}
\textit{Acknowledgments}---This work is supported by the National Natural Science Foundation of China (12672216, 12322207, T2293720/T2293722). We thank Prof. Xiang Yu of Dongguan University of Technology for helpful discussions.
\vspace{\baselineskip}

W. K., Y. Z., and C. L. contributed equally to this work. X. L. and Z. W. conceived the idea and supervised the project.

\bibliography{}

% PRL specialist material follows the references and is headed by appendices.
\onecolumngrid
%\clearpage
\begin{center}
 {\large\bfseries Appendix}
\end{center}
\twocolumngrid
\appendix
\setcounter{secnumdepth}{1}

% BEGIN EMBEDDED APPENDICES
\section{Constitutive reduction and nonconvex relaxed energy}
\label{app:relaxation}

Mechanical stretches are measured relative to the stress-free configuration at the testing uniform temperature \(T\). The calibrated free-energy density with incompressible constraint is
\begin{equation}
\tag{A1}\label{eq:appA1}
\begin{aligned}
\Psi_T={}&\frac{\mu(T)}{2}(\mathrm{tr}\,\mathbf C_N-3)
+\frac{\mu(T)\alpha}{2}\lambda^2\sin^2\theta+\frac{\mu(T)k_1}{2}(I_4^N-1)^2\\
&+\frac{A_Q}{2}(T-T^*)Q^2-\frac{B_Q}{3}Q^3+\frac{C_Q}{4}Q^4,\\
\end{aligned}
\end{equation}
where $\mathbf C_N=\mathbf F_N^{\mathrm T}\mathbf F_N$, $\mathbf F_N=\bar{\boldsymbol\ell}^{-1/2}\mathbf F_m
\bar{\boldsymbol\ell}_0^{1/2}$, $\bar{\boldsymbol\ell}=(\det\boldsymbol\ell)^{-1/3}\boldsymbol\ell$, $\boldsymbol\ell=(1-Q)\mathbf I+3Q\,\bm{d}\otimes\bm{d}$, and $I_4^N=\mathbf C_N:(\bm{d}_0\otimes\bm{d}_0)$, with \(\bar{\boldsymbol{\ell}}_0
=\bar{\boldsymbol{\ell}}(Q_0,\bm d_0)\) defined by the reference order \(Q_0\) and imprinted director \(\bm{d}_0\) at prescribed temperature $T$. The coefficients \(\alpha\) and \(k_1\) characterize the semi-soft and anisotropic penalties, respectively, while \(A_Q\), \(B_Q\), \(C_Q\), and \(T^*\) specify the nematic-isotropic nematic ordering free energy. 

For homogeneous extension along $\mathbf e_z$ transverse to the imprinted director $\bm{d}_0 =\mathbf e_y$, the deformation and current director are $\mathbf F_m=\operatorname{diag}[\lambda_x,(\lambda\lambda_x)^{-1},\lambda]$ and $\bm{d}(\theta)=\cos\theta\,\mathbf e_y+\sin\theta\,\mathbf e_z$. With $\mathbf m=(\lambda_x,\theta,Q)$ and $\hat f(\lambda,\mathbf m;T)=\Psi_T(\mathbf F_m,\bm{d}(\theta),Q;T)$, relaxing the internal variables at fixed lateral traction gives the homogeneous energy density
\begin{equation}
\tag{A2}\label{eq:appA2}
\Phi(\lambda;T)=\min_{\mathbf m}\hat f(\lambda,\mathbf m;T).
\end{equation}
The incompressibility constraint is enforced exactly by the kinematic parametrization of $\mathbf F_m$. Minimizing with \(\widehat f_{\mathbf m}=0\) and positive-definite \(\widehat f_{\mathbf m\mathbf m}\), differentiation of the stationarity condition gives the relaxed axial tangent
\begin{equation}
\tag{A3}\label{eq:appA3}
\Phi''=\hat f_{\lambda\lambda}
-\hat f_{\lambda\mathbf m}
(\hat f_{\mathbf m\mathbf m})^{-1}
\hat f_{\mathbf m\lambda},
\end{equation}
where the derivatives are evaluated in independent coordinates of the constrained relaxation problem. The first term is the frozen modulus, and the second is nonpositive when $\hat f_{\mathbf m\mathbf m}>0$. Coupled relaxation of the director angle $\theta$ and nematic order parameter $Q$ can reduce the axial tangent modulus below zero over a finite stretch interval, producing the nonconvex relaxed energy that permits coexistence of low- and high-stretch states at fixed mean stretch. Material parameters and calibration details are provided in the SM~\cite{SM}.

\section{Gibbs driving force and reverse-transformation resistance}
\label{app:coexistence}
At prescribed temperature, two locally stable homogeneous states $\lambda_\pm(P,T)$ satisfy $\Phi'(\lambda_\pm,T)=P$ and $\Phi''(\lambda_\pm,T)>0$. Their Gibbs-energy difference is
\begin{equation}
\Delta g(P,T)
=g(\lambda_+;P,T)-g(\lambda_-;P,T),
\end{equation}
where $g(\lambda;P,T)=\Phi(\lambda,T)-P\lambda$. Along either stable homogeneous state, differentiation with respect to $P$ at fixed temperature gives
\begin{equation}
\begin{aligned}
\frac{\mathrm{d}g(\lambda_\pm(P,T);P,T)}
{\mathrm{d}P}
=
\left[\Phi'(\lambda_\pm,T)-P\right]
\frac{\partial\lambda_\pm}{\partial P}
-\lambda_\pm
=-\lambda_\pm,
\end{aligned}
\label{eq:app_gibbs_derivative}
\end{equation}
where the equilibrium condition eliminates the implicit stretch dependence. Consequently,
\begin{equation}
\frac{\mathrm{d}\Delta g}{\mathrm{d}P}
=-\left[\lambda_+(P,T)-\lambda_-(P,T)\right]
=-\Delta\lambda(P,T).
\label{eq:app_energy_derivative}
\end{equation}
Since $\Delta g(P_M^{\mathrm{eff}},T)=0$, integration from the loading-calibrated Maxwell reference to the measured unloading stress $P_U(T)$ yields
\begin{equation}
\Delta g_c(T)
=\Delta g(P_U(T),T)
=\int_{P_U(T)}^{P_M^{\mathrm{eff}}(T)}
\Delta\lambda(p,T)\,\mathrm{d}p.
\label{eq:resistance}
\end{equation}
If $\Delta\lambda(p,T)$ varies weakly over this interval, it can be approximated by $\Delta\lambda_M(T)=\Delta\lambda(P_M^{\mathrm{eff}}(T),T)$, yielding Eq.~\eqref{eq:plateau_gap} of the main text.

\section{Gradient regularization and interfacial energy}
\label{app:gradient}

At fixed temperature, the geometry in Eq.~\eqref{eq:gradient} refers to the stress-free thermal configuration, and
$D_{\rm eff}^r(T)>0$ is constant across the interface. Stationarity subject to $\int_0^{L_g}\lambda\,\mathrm dZ=L_g\bar\lambda$ gives
\begin{equation}
D_{\rm eff}^r\lambda_{,ZZ}=A[\Phi'(\lambda,T)-P],
\label{eq:appC_EL}
\end{equation}
Here $P$ enforces the mean constraint. Free endpoint stretches and the absence of end energy give the natural boundary conditions $\lambda_{,Z}(0)=\lambda_{,Z}(L_g)=0$. The common tangent to the loading-calibrated energy defines the effective Maxwell stress $P_M^{\mathrm{eff}}(T)$ and the loading coexistence stretches $\lambda_\pm^{\mathrm 
M}(T) =\lambda_\pm(P_M^{\mathrm{eff}}(T),T)$. Integrating this equation and imposing zero gradients in both far fields of an isolated interface requires equal Gibbs potentials at $P=P_M^{\rm eff}$. Hence
\begin{equation}
\frac{D_{\rm eff}^r}{2A}\lambda_{,Z}^2
=\Delta\Phi(\lambda,T),
\label{eq:appC_first}
\end{equation}
Here, $\Delta\Phi(\lambda,T)
=\Phi(\lambda,T)-\Phi(\lambda_-^M,T)-P_M^{\rm eff}(\lambda-\lambda_-^M)$ is the excess above the common tangent connecting the loading coexistence stretches $\lambda_-^M$ and $\lambda_+^M$, and vanishes at both endpoints. The loading-calibrated Maxwell construction defines the reference coexistence stretches \(\lambda_\pm^M(T)=\lambda_\pm(P_M^{\mathrm{eff}}(T),T)\). In the calculations below, we denote these loading-reference states by \(\lambda_\pm^L\equiv\lambda_\pm^M\). They define the common-tangent barrier and its integration limits for both loading and unloading reference interfaces, with the respective profile fits determining \(D_{\mathrm{eff}}^L\) and \(D_{\mathrm{eff}}^U\). 

The interfacial excess free energy per unit reference area is
\begin{equation}
\begin{aligned}
\gamma_{\mathrm{int}}^r(T)
=2\int_{-\infty}^{\infty}\Delta\Phi\,\mathrm dZ.
\end{aligned}
\label{eq:appC_gamma}
\end{equation}
Orienting the profile from low to high stretch and changing variables using $\mathrm dZ=
\sqrt{D_{\mathrm{eff}}^r/(2A\Delta\Phi)}\,\mathrm d\lambda$ gives
\begin{equation}
\gamma_{\mathrm{int}}^r(T)
=
\frac{1}{A}
\int_{\lambda_-^M}^{\lambda_+^M}
\sqrt{
2AD_{\mathrm{eff}}^r(T)\Delta\Phi(\lambda,T)
}\,\mathrm d\lambda.
\label{eq:interface_energy}
\end{equation}
Both paths use the same loading-reference barrier and endpoints, with their respective fitted coefficients $D_{\mathrm{eff}}^r$ identified from DIC profiles as described in the SM~\cite{SM}.

\section{Finite-length branch selection and tensile accessibility}
\label{app:finite-length}

We compare the energies in both homogeneous and localized configurations using the same thermal reference geometry, $D_{\rm eff}^L(T)$, and boundary conditions. The localized profile $\lambda_i$ satisfies Eq.~\eqref{eq:appC_EL} and the prescribed mean stretch. Define $\Delta f=(\mathcal F_{\rm nec}-\mathcal F_{\rm hom})/(AL_g)$. Along smooth stationary branches, integration by parts, the natural boundary conditions, and the mean constraint give
\begin{equation}
\begin{aligned}
\frac{\partial\Delta f}{\partial\bar\lambda}=P_{\rm nec}-P_{\rm hom}.
\end{aligned}
\label{eq:appD_envelope}
\end{equation}
At the equal-energy crossing $\Delta f(\bar\lambda_u,T)=0$, define the stresses for the homogeneous and localized configurations in unloading as $P_{\alpha}^u=P_\alpha(\bar\lambda_u,T)$. Then
\begin{equation}
\begin{aligned}
\Delta f(\bar\lambda_u+\delta\bar\lambda,T)
={}&(P_{\rm nec}^u-P_{\rm hom}^u)\delta\bar\lambda+o(\delta\bar\lambda).
\end{aligned}
\label{eq:appD_crossing}
\end{equation}
Thus $P_{\rm nec}^u>P_{\rm hom}^u$ makes the localized branch energetically preferred upon contraction through a simple crossing. Equal energies need not imply equal stresses. The stability must be assessed separately, as stationarity and relative energy do not establish stability. 

Consider ideal quasistatic unloading starting above $\bar\lambda_u$ on the homogeneous branch. Assume this branch remains stable and accessible, its stress decreases monotonically as $\bar\lambda$ decreases, and switching occurs when localization becomes energetically preferred. If unloading stops at its first zero-force state, $P_{\rm hom}^u>0$ places the crossing within the tensile unloading path. The boundary is defined, when a zero crossing exists, by
\begin{equation}
P_{\rm hom}^u(T_c)
=\Phi'(\bar\lambda_u(T_c),T_c)=0.
\label{eq:appD_threshold}
\end{equation}
For $P_{\rm hom}^u<0$, the crossing lies beyond that zero-force limit. This defines energetic access to reverse localization. For a nonzero stopping stress, the boundary becomes $P_{\rm hom}^u(T)=P_{\rm stop}(T)$. Complete recovery additionally requires a stable, accessible localized path during further unloading and sufficient driving force against resistance to reverse boundary motion. The numerical solution of the finite-length problem
and the extraction of the equal-energy crossings and corresponding branch stresses are described in the SM~\cite{SM}.
% END EMBEDDED APPENDICES
\end{document}